\documentclass[prb,,twocolumn,showpacs,aps,floats,floatfix]{revtex4}

\usepackage{epsfig}
\usepackage{epstopdf}
\begin{document}

\title{Thermodynamics of  nano-spheres encapsulated in virus capsids}

\author{Antonio \v{S}iber}
\email{asiber@ifs.hr}
\affiliation{Institute of Physics, Bijeni\v{c}ka cesta 46, 10000 Zagreb, Croatia}

\author{Roya Zandi}
\email{roya.zandi@ucr.edu}
\affiliation{Department of Physics and Astronomy, University of California, Riverside, California 92521, USA}

\author{Rudolf Podgornik}
\email{rudolf.podgornik@fmf.uni-lj.si}
\affiliation{Department of Theoretical Physics, Jo\v{z}ef Stefan Institute, SI-1000 Ljubljana, Slovenia}
\affiliation{Institute of Biophysics, School of Medicine and Department of Physics, University of Ljubljana, SI-1000 Ljubljana, Slovenia}
\affiliation{Laboratory of Physical and Structural Biology, Eunice Kennedy Shriver National Institute of Child Health and Human Development, NIH, Bethesda, Maryland 20892-0924, USA}

\begin{abstract}
We investigate the thermodynamics of complexation of functionalized charged nano-spheres with viral proteins. 
The physics of this problem is governed by electrostatic interaction between the proteins and the 
nano-sphere cores (screened by salt ions), but also by configurational degrees of freedom of the charged protein N-tails. 
We approach the problem by constructing an appropriate complexation free energy functional. On the 
basis of both numerical and analytical studies of this functional we construct the phase diagram for 
the assembly which contains the information on the assembled structures that appear in the 
thermodynamical equilibrium, depending on the size and surface charge density of the nano-sphere cores. We show that both the 
nano-sphere core charge as well as its radius determine the size of the capsid that forms around the core.
\end{abstract}
\date{\today}
\maketitle

\section{Introduction}

Viruses have optimized the feat of packaging genome molecules and delivering them to the appropriate cells.  In its simplest form, a virus consists of a rigid 
protein shell (capsid) that surrounds and protects the genetic material (either RNA or DNA) from chemical and physical assaults \cite{Flint}. A viral capsid 
contains several copies of either one type of protein or of a few slightly different kinds. A number of {\it in vitro} self-assembly experiments reveal that 
the protein subunits of many RNA viruses can assemble spontaneously not only around their own genome but the genomes derived from other viruses and various non-viral 
anionic polymers. All these features in addition to their extraordinarly highly symmetric shape and monodisperse size distributions make viruses ideal structures 
for gene therapy, drug delivery and various nanotechnology and material science applications 
\cite{Group_Douglas_Young,Lee_Mao_dots,Lewis_virus_vascular_imaging,Comellas-Aragones_CCMV_nanoreactors}. To this end, the number of experiments and theoretical research 
aimed at understanding the physical basis of assembly of viruses and the factors influencing the structure and size of viral capsids is amazingly soaring 
\cite{Barklis,Benjamin,Ganser,Zlotnick,Michel,Hu,sibrudi08,Rapaport_viral_capsid_growth,Glotzer_elongated_viruses,Brooks_capsid_polymorphism,Zandi,Twarock_assembly_papovaviridae,sibrudi1,Hagan,Levandovsky}.

The majority of viral capsids have either spherical or elongated structures.  Most spherical viruses have structures with icosahedral symmetry and contain $60 {\cal T}$ protein subunits 
where ${\cal T}$ is the structural index of viral shells and is determined from the relation ${\cal T}=h^2+k^2+kh$ with $h$ and $k$ nonnegative integers \cite{Caspar}. While the capsid 
protein of some viruses can form only one structure with a specific size, the capsid protein of many others are more flexible and adopt various structures with different sizes. 
For example, tobacco mosaic virus (TMV) capsid proteins assemble into tubular structures regardless of the shape and size of their cargo \cite{Klug}. This indicates that the 
shape of the virus is solely dictated by the intrinsic property of protein subunits. In the other end of the spectrum, many experiments show that the capsid proteins of cowpea 
cholorotic mosaic virus (CCMV), a spherical RNA virus, are able to form capsid of various size and shapes \cite{Bancroft,Adolph,Lavelle}.

Over 35 years ago, Adolph and Butler\cite{Adolph} and more recently Lavelle \textit{et al.} \cite{Lavelle} performed a series of {\it in vitro} experiments with CCMV capsid 
proteins in the absence of RNA and found that depending on the pH and ionic strength several different structures assembled. A notable feature of the constructed shape "phase" 
diagram based on these experiments is the change from an icosahedral ${\cal T}=3$ structure to a cylindrical shape upon a decrease in ionic strength and an increase in pH 
revealing the important role of electrostatic interaction on the size and shape of empty viral shells. 

There have been a number of different experiments and theoretical studies \cite{Dragnea1,Hagan_assembly_templating,ZandiSch,Bruinsma,VdSchootBruinsma,Belyi,sibrudi2,Dragnea2} to 
investigate the impact on the structure of capsids of the shape and length of genome. To explore the effect of cargo on the morphology of viral capsid, Zlotnick and coworkers 
examined the assembly of capsid proteins of CCMV around heterogeneous DNA longer than 500 base pairs and found that tubular structures spontaneously formed \cite{Zlotnick}. Note 
that CCMV has at ${\cal T}=3$ icosahedral structure in its native form.  

Quite remarkably, almost half a century ago, Bancroft and coworkers (see e.g. Ref. \onlinecite{Bancroft}) demonstrated the important role of stoichiometry ratio of capsid proteins 
and genome in the structure of viral shells. According to their experiments CCMV capsid proteins could encapsidate TMV RNAs which is about 6000 nucleotides in viral particles of 
various sizes. Depending on the ratio of RNA-protein concentrations, ${\cal T}=3$, ${\cal T}=4$ or ${\cal T}=7$ structures can form. 

More recent experiments with PSS (polystyrene sulfonate, a highly flexible polyelectrolyte chain) and CCMV capsid proteins indicate the significance of length of genome in that 
the size of CCMV viral shells encapsidating PSS vary from 22 nm to 27 nm when the molecular weight of PSS varies from 400KDa to 3.4 MDa \cite{Hu}. The impact of the size of cargo 
on the diameter of capsid is quite transparent in the experiments of Dragnea and coworkers in which they found that the capsid proteins of Bromo Mosaic Virus, another spherical 
plant virus, are able to package the functionalized gold nano-spherical particles of diameters of 6, 9 and 12 nm to form virus like particles (VLP) with ${\cal T}=1$, ${\cal T}=2$, 
or ${\cal T}=3$ structures, respectively \cite{Dragnea1}.

All the aforementioned experiments focus on the important role of size and structure of genome and stoichiometric ratio of genome to protein in determining the structure of viral 
shells. However, several experimental and theoretical studies reveal that electrostatic interaction is the driving force for the assembly of capsid proteins around anionic cargos, 
and thus it is crucial to study the impact of cargo charge density on the structure of capsids. In fact, a careful study of several single stranded RNA viruses show that there is 
a linear relation between the number of charges on the capsid inner surface and on their genome \cite{Belyi}.  An important question, then, naturally arises: Could we change the 
size of a capsid by changing the net charge of its cargo?  More specifically, in the experiments of Dragnea and coworkers with nano-spheres, does a ${\cal T}=3$ structure form, 
if one increases the charge density of the 9 nm cores which normally form "pseudo" ${\cal T}=2$ structures? 

In this paper, we investigate the interplay between the charge density and size of nano-cargos in virus assembly. Similar to the experiments of Dragnea and coworkers, we consider 
negatively charged nano-spheres which interact with positively charged capsid inner surface, under physiological condition and find that in addition to the diameter of the 
encapsidated nano-spheres, the total net charges on cargos have a significant impact on the size of viral capsids. 

An important feature of several RNA viruses, including CCMV and BMV mentioned above, is the presence of cationic polypeptide chains that form the N-termini 
of the capsid protein.  Rich in basic amino acids, there is a total of thousands of charges on the N-terminal tails which extend into the capsid interior and are 
responsible for the absorption of RNA to the inner capsid surface.  Very recent in vitro studies of Aniagyei \textit{et al.} reveal that a mutant of CCMV coat proteins lacking 
most of the N-terminal domain, $N\Delta 34$, assemble around negatively charged 12 nm spherical cores to form ${\cal T}=2$ structures \cite{Dragnea2}. Note that native CCMV 
proteins form a ${\cal T}=3$ structure around 12 nm spherical cores.  Our calculations also show that the N-terminal arms can have a major impact on the virion structure and 
(as shown in Fig. 3 and 4), they can significantly modify the free energy landscape of viral structures. One also has to consider that the deletion of the N-terminal tails might change 
the preferred angle between the protein subunits, an effect which is not taken into account in the present study.  According to our studies, depending on the cargo charge 
density and the presence of N-terminal tails it might be advantageous for capsid proteins to form relatively smaller or bigger shells compared to their native structures. The effect 
of N-terminal on the free energy of viral capsids has been investigated 
previously \cite{HaganJCP}. Here we take another approach that enables us to study the energetics of 
complexation of proteins and core in more details. Our emphasis is also on different aspects of the 
assembly, in particular the formation of differently sized structures depending on the conditions.

The outline of the paper is as follows. In Section II, we present our model to calculate the free energy between capsid inner surface and a rigid sphere including the 
interaction of positively charged N-terminal tails with the spherical cargo.  In Sec. III, we present our numerical results and in Sec. IV we discuss our findings and 
their implications, and summarize our conclusion.

\section{Theoretical description of energetics and thermodynamics of the assembly}
\label{sec:theory}

Here, we consider encapsidation of charged nano-particles (whose number is $n_c$) within the virus capsid and the dependence of 
the formation free energy on the parameters describing the system.  We assume that the solution consists of monovalent salt 
(of bulk concentration $c_0$), dissolved protein monomers (or, more generally, basic protein subunits, which may be e.g. protein dimers as is the 
case for hepatitis B virus - their number is assumed to be $n_p$) and spherical cores that are perfectly monodisperse with respect to radius 
$R_1$ and surface charge density $\sigma_1$. All the particles are dissolved in a medium whose dielectric constant is $\epsilon_0 \epsilon_r$ (we 
shall take $\epsilon_r=80$, i.e. water). An assembly problem of this type involves many parameters, including the concentrations of cores ($n_c / V$)
and proteins ($n_p / V$, where $V$ is the volume of the solution). These two parameters importantly influence the assembly phase diagram, 
in addition to the energy of the assembled complex, which is a fairly complicated quantity in itself. One can expect 
formation of variously sized protein capsids (described by different Caspar-Klug ${\cal T}$ numbers) around the core, 
depending both on the properties of the core but also on particle concentrations ($n_c/V$ and $n_p/V$). We shall first, 
however, examine the energetics of the formed complex (capsid + core). 

\subsection{Energetics of the assembled complex}

\begin{figure}[h]
\centerline{
\epsfig {file=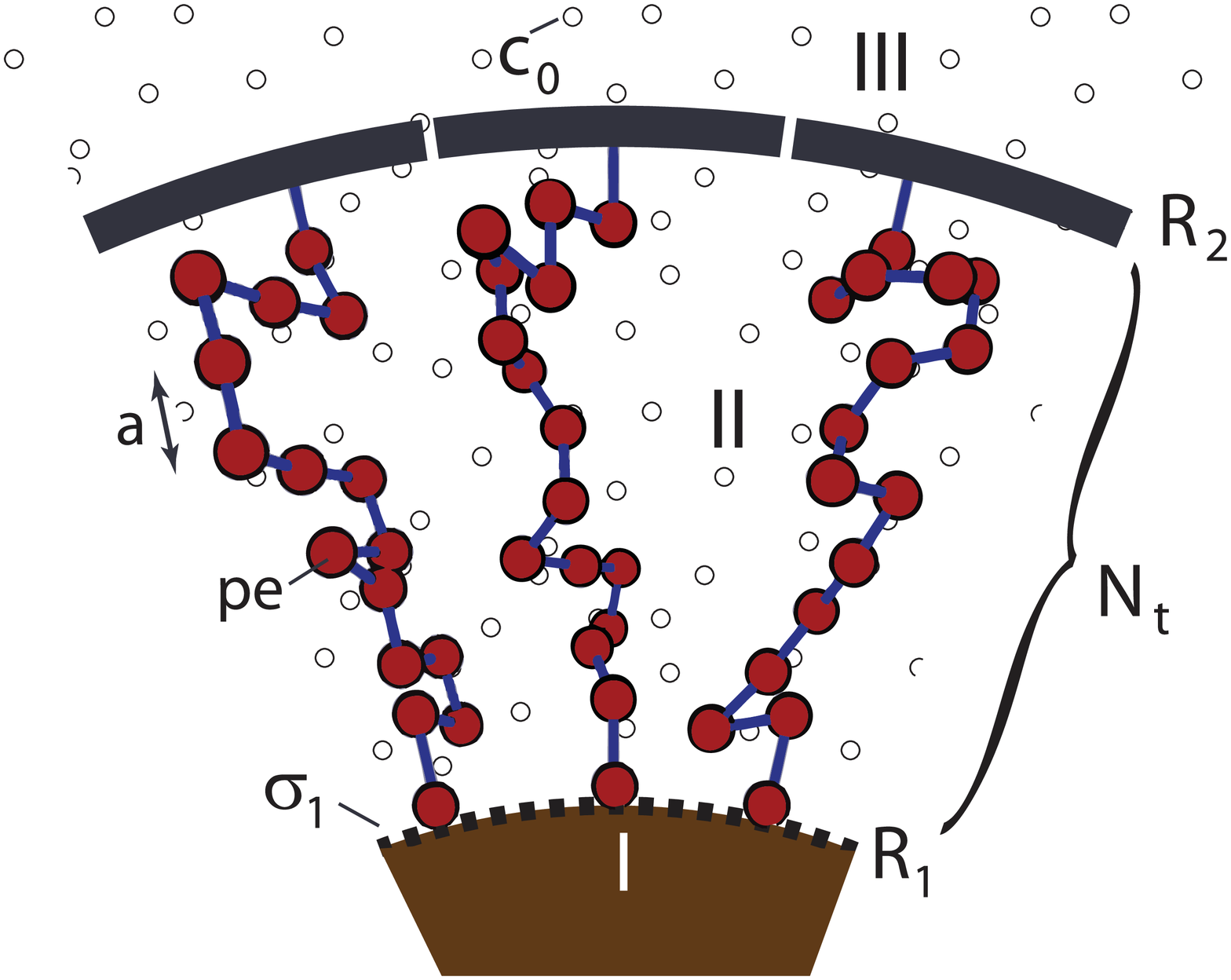,width=8.5cm}
}
\caption{Illustration of the portion of the spherical complex of proteins with flexible tails and a charged core. The parameters discussed in the text are indicated. 
The configuration of the tails corresponds to the case when the core is poorly charged.}
\label{fig:ilustracija}
\end{figure}

In general, we shall assume that the protein charge is distributed along the flexible N-tails. We treat the tails as generic polyelectrolytes 
with intermonomer bondlength $a$, partial charge $p$ per monomer, and non-electrostatic excluded volume interactions characterized 
by the excluded volume $v$. Our approach is quite similar to that exposed in Ref. \onlinecite{sibrudi2}. The free energy 
of the complex in the subspace of fixed total number of polyelectrolyte monomers, $N$, can be calculated from
\begin{equation}
F=\int f(r) d^3r - \mu \left ( \int d^3 r |\Psi ({\bf r})|^2 - N \right ), 
\end{equation}
where $\mu$ is the Lagrange multiplier enforcing the condition of fixed number of monomers, and 
\begin{widetext}
\begin{eqnarray}
f(r) &=& k_B T \left[ \frac{a^2}{6} |\nabla \Psi(r)|^2 + \frac{v}{2}\Psi(r)^4 \right ]
+ e c^{+}(r) \Phi(r) - e c^{-}(r) \Phi(r) +p e |\Psi(r)|^2 \Phi(r) - \frac{\epsilon_0 \epsilon}{2} |\nabla \Phi(r)|^2 \nonumber \\
&+& \sum_{i= \pm} \left \{ 
k_B T \left [ 
c^{i}(r) \ln c^{i}(r) - c^{i}(r) - \left( 
c_0^{i} \ln c_{0}^{i} - c_{0}^{i}
\right)
\right ] \right.
- \left .\mu^{i} \left[ 
c^{i}(r) - c_{0}^{i}
\right]
\right \}.
\label{eq:fions}
\end{eqnarray}
\end{widetext}
The free energy is a functional of the monomer density field of the polyelectrolyte chains ($\Psi^2 ({\bf r})$) and the mean electrostatic potential ($\Phi(r)$). 
The complex free energy also depends on the salt concentration fields ($c^{\pm}(r)$) whose chemical potentials are denoted by $\mu^{\pm}$.
One can assume that in addition to the charge located on mobile protein tails, there is also a static density of charge, $\rho_p(r)$, which could 
in principle be located on the immobile parts of the capsid proteins (outside tails). In the further calculations, we shall neglect the fixed charge on 
the capsid and assume that the static charge in the system resides exclusively on the core surface, so that 
\begin{equation}
\rho_p({\bf r}) = \sigma_1 \delta(r - R_1),
\end{equation}
where $\sigma_1$ is the charge density at the core surface. The contribution of such localized charges to the electrostatic part of the 
free energy can be easily separated (if required) from the functional in the form of the boundary term as it was done in Ref. \onlinecite{sibrudi1}. See 
Fig. \ref{fig:ilustracija} for the illustration of the system that we consider and the relevant parameters characterizing it.

The variation of the free energy functional with 
respect to fields $\Psi$, $\Phi$, and $c^{\pm}$ yields two coupled partial differential equations. 
The first one is the generalized polyelectrolyte Poisson-Boltzmann equation, 
\begin{equation}
\epsilon_0 \epsilon \nabla ^2 \Phi(r) = 2 e c_0 \sinh [\beta e \Phi(r) ] - e p \Psi(r)^2 + \rho_p(r).
\label{eq:PBE}
\end{equation}
The second equation is the Edwards equation,
\begin{equation}
\frac{a^2}{6} \nabla^2 \Psi(r) = v \Psi(r)^3 + p e \beta \Phi(r) \Psi(r) - \mu \Psi(r).
\label{eq:polyel}
\end{equation}
The Lagrange multiplier ($\mu$) enforces the conservation of the total number of polyelectrolyte monomers,
\begin{equation}
\int d^3 r \Psi^2 ({\bf r}) = N_p N_t = N,
\label{eq:normalizacija}
\end{equation}
where $N_t$ is the number of monomers in a particular tail, and $N_p$ is, as before, the number of 
proteins (or, in general, protein subunits) in the complex. The electrostatic boundary conditions are the same as in the standard PB theory. In 
particular, there is a boundary condition at $r=R_1$, reflecting the finite charge density, $\sigma_1$ at the surface of the core.
There are two additional conditions that need to be specified for the density field $\Psi(r)$. We simply take the core 
to be impenetrable and thus presume that 
\begin{equation}
\Psi(R_1) = 0.
\label{eq:impen1}
\end{equation}
On the inner part of the capsid one could assume that
\begin{equation}
\Psi(R_2) = \sqrt{\frac{N_p}{4 \pi a R_2^2}},
\label{eq:anchor}
\end{equation}
which means that there must be some density of the tails at the capsid wall since they are fixed 
or grafted. This is the only way that grafting enters our calculation.  This condition signifies that in 
the volume $4 \pi a R_2^2$  of the spherical shell, touching the interior of the capsid with its larger radius, 
there should be {\em one} monomer per each of the 
tails in the structure. In the continuum limit, this means that the polyelectrolyte anchors to the capsid 
perpendicularly (see Fig. \ref{fig:ilustracija}).

The procedure presented thus far is based on the mean-field description of the electrostatics of the problem and 
ground-state dominance ansatz for the polyelectrolyte field, see Ref. \onlinecite{sibrudi2} for details. Furthermore, 
the finite extensibility of the polyelectrolyte is not taken into account. While that was of no essential importance 
for the problem studied in Ref. \onlinecite{sibrudi2}, it may become of importance in our case, since the protein 
tails are assumed to be grafted to the capsid. To take the geometrical constraints regarding the polyelectrolyte density 
field into account, we shall now derive an alternative set of equations that we will obtain 
from a constrained variation of a free energy functional. Instead of varying $F_{complex}$ over the space of all functions 
$\Psi({\bf r})$, we shall vary it over the {\em constrained} set of functions, i.e. polyelectrolyte amplitudes that can be represented as 
\begin{equation}
\Psi(r) = \Psi_{s}(r) + u^2(r),
\label{eq:defin1}
\end{equation}
where $u(r)$ is a real function and 
\begin{equation}
\Psi_{s}^2(r) = \frac{N_p}{4 \pi a r^2}.
\label{eq:maxstretch}
\end{equation}
The above equation describes the maximally extended polyelectrolyte density, so that no smaller values of $\Psi(r)$ are possible without 
enlarging the monomer-monomer separation distances $a$ (stretching). Varying $F_{complex}$ over $u({\bf r})$, we obtain the new set of 
Euler-Lagrange differential equations for $\Phi(r)$ and $u(r)$ as
\begin{equation}
\epsilon_0 \epsilon \nabla ^2 \Phi(r) = 2 e c_0 \sinh [\beta e \Phi(r) ] - e p [\Psi_s(r)+u^2(r)]^2, 
\label{eq:PBnew}
\end{equation}
and
\begin{equation}
\frac{a^2}{6} {\cal L}_{\Psi_s} [u(r)] = s (r)
\label{eq:Edwardsnew}
\end{equation}
where ${\cal L}_{\Psi_s} (u)$ is a differential operator given as
\begin{equation}
{\cal L}_{\Psi_s} (u) =  \nabla u \nabla \Psi_s + 2 [ u^2 \nabla^2 u + 2 u (\nabla u)^2].
\end{equation}
We have used here $\nabla^2 \Psi_s=0$. Function $s(r)$ is given as
\begin{eqnarray}
s(r) &=& v (\Psi_s^3u + 3 \Psi_s^2 u^3 + 3 \Psi_s u^5 + u^7 )  
\nonumber \\
&+& \beta e \Phi (\Psi_s u + u^3 ) - 2 u \mu (\Psi_s + u^2).
\end{eqnarray}
We still have to specify the boundary conditions. Looking at Eq. (\ref{eq:defin1}), one sees that 
we can no longer put $\Psi(R_1)=0$, so we will be necessarily stuck with finite density of 
monomers at the core radius. We choose
\begin{equation}
u(R_1) = \delta, \; u(R_2) = \delta, \; \delta \rightarrow 0
\label{eq:ubound}
\end{equation}
as the appropriate boundary conditions implying that the N-tails are {\em normal} to the surface of the inner core as well as outer 
capsid wall at $R_1$ and $R_2$. In other words we assume that the chains have no overhangs at $R_1$ and $R_2$, so that they touch 
both the core and the capsid only once and perpendicularly. Note that Eq. (\ref{eq:anchor}) is automatically satisfied with such a choice. 

The complex free energy that we have constructed thus far accounts approximately for electrostatic energy of the system, the 
entropic and excluded volume effects of the confined polyelectrolyte (protein tails), but also for entropic contributions of 
the salt ions (on the mean-field level). It does not contain, however, the attractive component (non-electrostatic) of protein-protein 
interactions. These interactions consist of hydrophobic and van der Waals contributions \cite{Israela} and we denote 
their value per protein in the complex as $\overline{f}_{p,hydro}$. It is this part of the energy that keeps the dominantly 
positively charged {\em empty} capsids together \cite{SchKeg,sibrudi1}.

\subsection{Thermodynamics of the assembly}
\label{subs:thermod}

The parameter space for assembly 
that we are interested needs to be at least four-dimensional ($R_1$, $\sigma_1$, $n_c / V$, and $n_p / V$), presuming 
that the properties of the capsid proteins are kept fixed. Analysis of assembly in such a high-dimensional parameter space would 
be highly involved. \cite{ZandiSch,SibMaj} It is of interest thus to try to extract the relevant information on the assembly {\em solely} from the complex free energy, as 
we have defined it. Such a procedure cannot be expected to be valid for all values of $n_c$ and $n_p$, but it should be of use when 
\begin{equation}
n_c \ll \frac{n_p}{N_p({\cal T}_{max})}, 
\label{eq:condit1}
\end{equation}
where ${\cal T}_{max}$ is the maximal capsid ${\cal T}$-number that can be possibly assembled in the experimental circumstances, and 
$N_p({\cal T}_{max})$ is the number of proteins (subunits) in a Caspar-Klug structure of that ${\cal T}$-number. Basically, Eq. 
(\ref{eq:condit1}) says that, concerning the ''final'' (assembled) state, there is no big 
difference with respect to the ${\cal T}$ number of the dominantly assembled structures. The final state will always 
consist of all cores complexed with certain number of proteins, and the rest will be, more or less the same 
number of proteins that may remain isolated in the solution or perhaps form empty capsids. There will be no free (uncomplexed) 
cores in the assembled state. 

Let us assume that prior to the assembly, the solution contains isolated charged cores and individual viral proteins 
(or protein dimers or whatever the basic subunit of the assembly might be) - this is the ''initial'' state. 
In the ''initial'' state, the free energy of the system is
\begin{equation}
F_i = n_p \overline{f}_p^i + n_c F_{core}^i,
\end{equation}
where $\overline{f}_p^i$ and $F_{core}^i$ are the free energies per isolated protein and core in the inital state, 
respectively. These may in principle contain also the translational entropy contributions. After the assembly, the free energy is 
(''final'' state) 
\begin{equation}
F_f = n_c F_{complex} + n_c N_p \overline{f}_{p,hydro} + (n_p - n_c N_p) \overline{f}_p^{f}.
\label{eq:finalf}
\end{equation}
The first term on the RHS of Eq. (\ref{eq:finalf}) is what we can calculate - this is the free energy of the complex (note, however, that we do not 
calculate the entropic term regarding the translational freedom of the assembled structures). The number of complexes is 
exactly $n_c$ since we assumed that {\em all} cores are complexed with the proteins. The second term 
on the RHS of Eq. (\ref{eq:finalf}) is the part of the attractive energy of proteins assembled in the complexes that is 
difficult to calculate. It contains the hydrophobic and van der Waals protein-protein interactions, and per protein, the 
corresponding free energy is $\overline{f}_{p,hydro}$. The third term on the RHS of Eq. (\ref{eq:finalf}) is the free energy 
of the proteins in the final state that are {\em not} assembled in the complexes with cores ($n_p - n_c N_p$ of them). 
They may, however, be assembled in 
empty capsids. Thus, in general, free energy per such protein in the final state ($\overline{f}_p^{f}$) is not the same as 
the free energy per {\em isolated, individual} protein in the initial state ($\overline{f}_p^i$). If the proteins that do not 
complex with the cores indeed form aggregates (e.g. empty capsids) then one can expect that $\overline{f}_p^{f} < \overline{f}_p^{i}$.

The system will proceed from state $i$ to state $f$ if $F_f < F_i$. We want to examine the quantity $F_f$ for complexes that have 
different ${\cal T}$ numbers, i.e. we want to find $F_f$ for several different final states $f$. For all of these states, the inital 
assembly state is the same, so for two different final states $f_1$ and $f_2$ we can directly compare the corresponding free energies 
$F_{f_1}$ and $F_{f_2}$ and if $F_{f_1}<F_{f_2}$ we can say that state $f_1$ is more likely to be realized in the thermodynamical 
equilibrium. It is plausible to assume that 
\begin{equation}
|F_{complex}| \gg N_p |\overline{f}_{p,hydro} - \overline{f}_p^{f}|,
\label{eq:assumption}
\end{equation}
and if this is indeed the case, then we can examine $F_{complex}$ for different complexes and compare them mutually (for given ''initial'' state, 
i.e. charge density and radius of the cores) so to judge about the 
thermodynamically preferred states. In order to be able to construct a phase diagram, i.e. to compare the free energies for 
{\em different} initial states, we shall construct the quantity 
\begin{equation}
\Delta F \equiv F_{complex} - F_{core}^i.
\label{eq:complexation}
\end{equation}
This quantity should contain the biggest part of the assembly free energy, assuming Eq. (\ref{eq:assumption}) holds. In the following, 
we shall term $\Delta F$ as the assembly free energy, and we shall refer to $F_{complex}$ (also denoted as $F$) as the complex free 
energy or the free energy of the complex.

\section{Numerical evaluation of the model}
\subsection{Tailless protein subunits}

As already discussed, a prominent feature of the virus proteins assembled in a capsid are the N-tails that protrude into the capsid interior. 
This feature is typical for many viruses and can influence both the energetics of the protein-genome assembly \cite{sibrudi2} 
and its speed. The tails are typically very positively charged, and they are thus expected to play a prominent role in 
the assembly of proteins with negatively charged cores. The possibility of spatial redistribution of the tails 
is also expected to influence the assembly, so that one can expect an interplay between the electrostatics of the tails and their 
configurational entropy. All of these effects are included in our free energy functional, at least approximately. 
In this section, we shall emphasize the electrostatic aspect of the capsid proteins and 
effectively neglect the tail positional degrees of freedom. In order to do so, we treat the N-tails as infinitely 
rigid with the vanishing intermonomer bondlength $a$. The problem then reduces to electrostatic interactions 
only since the entropy of the N-tails is quenched. Omitting the hydrophobic, vdW etc. energy of the capsid proteins, 
one is left with a purely electrostatic part of the total free energy. 

As a first approximation one can smear the charge of the $N_p$ proteins uniformly over the external sphere of radius 
$R_2$, so that its surface charge density is $\sigma_2$. The problem as described by this model system is stil not completely 
trivial as the Poisson-Boltzmann (PB) equation describing it is of course 
nonlinear \cite{sibrudi1}. However, some insight can be obtained by linearizing the PB equation 
and solving it in this approximation (Debye-H\"{u}ckel, DH). The details are elaborated in the 
Appendix and the final result for the electrostatic free energy is equation (\ref{eq:horrible}). One can further 
simplify it by using $\kappa R_1 \gg 1$, which is a condition typically met for the experiments done on viruses 
at physiological salt concentrations ($\sim$ 100 mM)\cite{SchKeg,sibrudi1}. In this case, 
\begin{eqnarray}
\lim _{\kappa R_1 \gg 1} F_{complex} &=& \frac{\pi}{\epsilon_0 \epsilon_r \kappa} \left \{
2 R_1^2 \sigma_1^2 \right. \nonumber \\
&+& 4 R_1 R_2 \sigma_1 \sigma_2 e^{\kappa(R_1-R_2)} + R_2^2 \sigma_2^2 e^{2 \kappa(R_1-R_2)} \nonumber \\
&+& \left. R_2^2 \sigma_2^2 \right \}.
\label{eq:fvirlim}
\end{eqnarray}
One should note here that the salt resides only in compartments III and II (see Fig. \ref{fig:ilustracija}), so 
there is no symmetry in the formula regarding $\sigma_1$, $\sigma_2$ and $R_1$ and $R_2$. The first term in Eq. 
(\ref{eq:fvirlim}) is the electrostatic self-energy of the core, and the third term is the self-energy of the 
protein shell. 
Note that the self-energy of the core 
has a prefactor of 2 with respect to the analogous term for the protein shell. This is due to the 
fact that the shell is screened by the salt both from the inside and the outside, which is not the 
case for the impenetrable core (note also that the dielectric constant of the core does not 
figure in the final equations). 

The second term in Eq. \ref{eq:fvirlim} is the electrostatic interaction free energy between the core and the 
protein shell. One easily sees that it will be minimized when $\sigma_1$ and $\sigma_2$ have different signs. We 
see that it decreases quickly as the distance between the shell and the core increases, i.e. it decreases as 
$\exp [-\kappa (R_2 - R_1)]$. 

\begin{figure}[h]
\centerline{
\epsfig {file=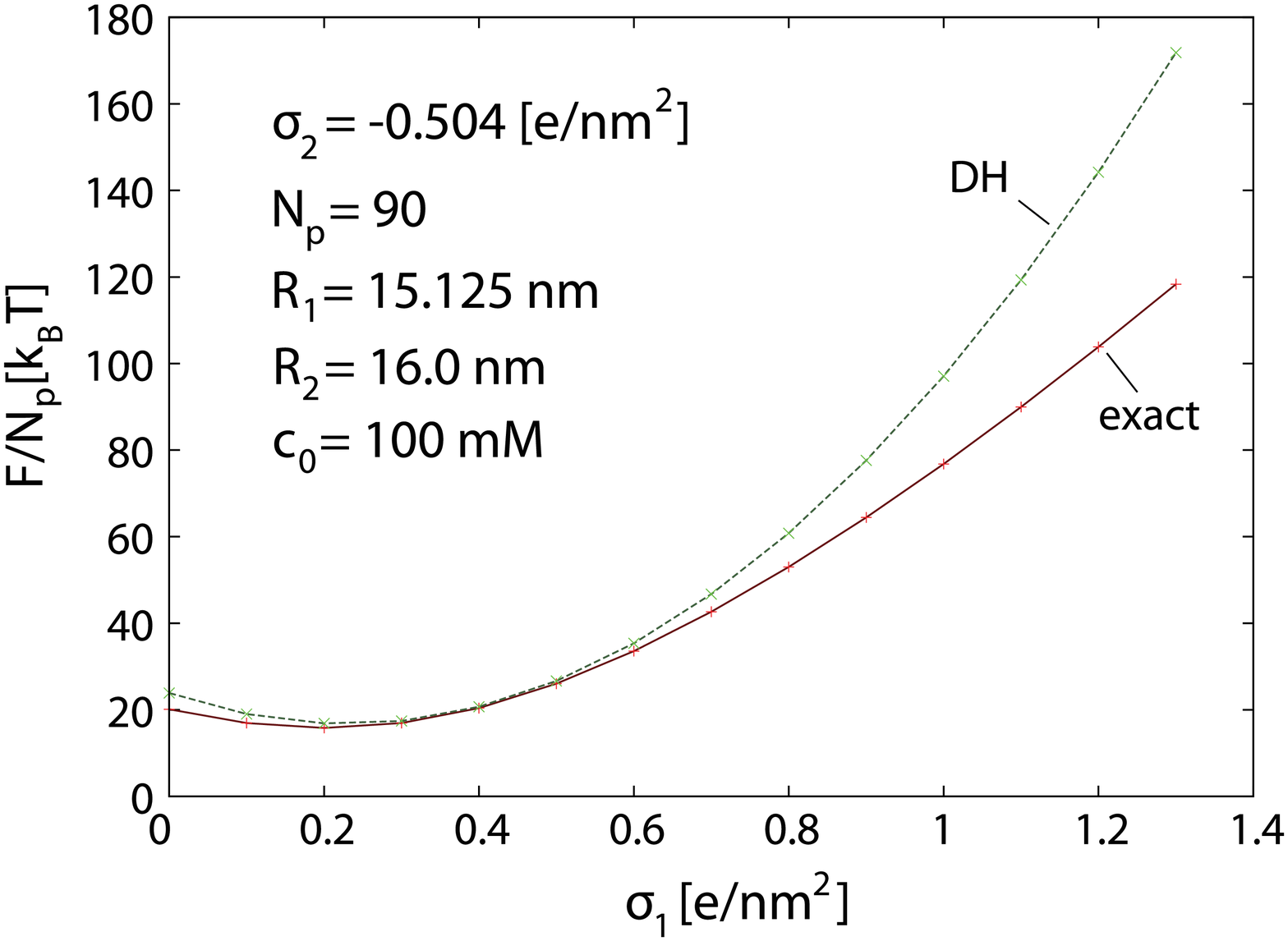,width=8.5cm}
}
\caption{Comparison of the numerically exact results with the analytic DH expression (\ref{eq:horrible}) for 
the free energy of the capsid with the charged core (complex free energy). The free energy per protein subunit is 
plotted as a function of the core surface charge density $\sigma_1$. The parameters of the calculation are denoted in the 
figure.}
\label{fig:figcomp1}
\end{figure}

Though the DH solution enables us to study the energetics of the assembly in more details, at least 
numerically one needs to check its validity vs. the complete non-linear Poisson-Boltzmann theory, see Ref. \onlinecite{sibrudi1}. By 
comparing it to the exact solution of the Poisson-Boltzmann equation one simultaneously checks the numerical results 
and the analytical formula for the DH solution. This comparison is shown in Fig. \ref{fig:figcomp1}.

We have chosen the capsid parameters to approximately represent the T=3 capsid of the BMV (brome mosaic virus) 
with capsid radius $R_2$ = 16 nm. The surface charge density of the capsid $\sigma_2$ = - 0.504 $e$/nm$^2$ was obtained by 
smearing the 18 charges per each of the $N_p=90$ dimeric tails over the capsid surface. Note that the absolute signs of 
the charge don't matter; what is important is that the charge on the core is of the opposite sign from the charge on the 
capsid. The core radius $R_1$ = 15.125 nm was chosen to be in the regime when the attractive interaction is not completely 
screened by salt - one can see this effect as a minimum in the free energy for some core charge density $\sigma_1$. This minimum 
disappears when the distance between the core and the capsid is larger than $\sim 1/\kappa$. In the 
chosen case, the minimum is at $\sigma_1 \sim 0.3$ $e$/nm$^2$.

As in the case studied in Ref. \onlinecite{sibrudi1}, the DH results always produce larger free energies than the 
exact Poisson - Boltzmann calculation. They are of course better when the potentials in the solution are small enough, so that 
the linearization approximation holds - this is the case for not too large surface charge densities, and 
it can be clearly seen that the DH approximation fails worse as $\sigma_1$ (or $\sigma_2$) increases, again 
in complete agreement with the results of Ref. \onlinecite{sibrudi1}.

Having established some confidence in the DH results, we can scrutinize them a bit more closely. If one 
takes $R_1 = R_2$ and $\sigma_1 = -\sigma_2$ in Eq. (\ref{eq:fvirlim}), one obtains the {\em absolute 
minimum} of the complex free energy in the whole $R_1, R_2, \sigma_1$ and $\sigma_2$ space. At that point in the 
parameter space the free energy is exactly {\em zero} which is the absolute minimum. This is due to the fact 
that the core charge exactly neutralizes the protein charge yielding effectively the uncharged shell 
($\sigma = \sigma_1 + \sigma_2 = 0$) of radius $R=R_1=R_2$. Note, however, that the assembly free energy is not 
bounded from below.

\begin{figure}[h]
\centerline{
\epsfig {file=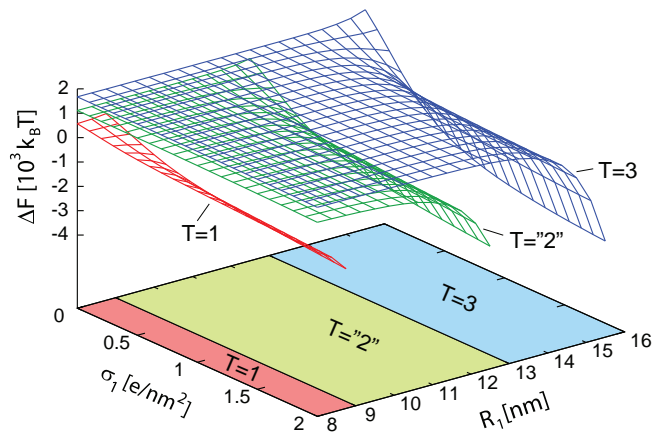,width=8.5cm}
}
\caption{Numerically exact results for the assembly free energy (Eq. \ref{eq:complexation}) assuming tailless protein 
subunits ($c_0 = 100$ mM). The free energies were plotted for capsids with three different ${\cal T}$-numbers as functions of the 
core surface charge density $\sigma_1$ and radius $R_1$. The number of 
capsomeres and corresponding capsid radii are $N_p$=30 (T=1), 60 (T=''2''), and 90 (T=3), and $R_2=9.24$ (T=1), 13.06 (T=''2''), 
and 16 nm (T=3).}
\label{fig:figexnotails}
\end{figure}

We now insert Eq. (\ref{eq:fvirlim}) in Eq. (\ref{eq:complexation}) and find
\begin{eqnarray}
\Delta F &=& \frac{\pi}{\epsilon_0 \epsilon_r \kappa} \left \{ 4 R_1 R_2 \sigma_1 \sigma_2 e^{\kappa(R_1-R_2)} \right. \nonumber \\
&+& \left. R_2^2 \sigma_2^2 e^{2 \kappa(R_1-R_2)} + R_2^2 \sigma_2^2 \right \}.
\label{eq:deltfDH}
\end{eqnarray}
Examining the assembly free energies (Eq. (\ref{eq:complexation})) of the variously sized ($R_1$) and charged ($\sigma_1$) cores 
with the proteins assembled in capsids of three different ${\cal T}$ number, we obtain the exact results for the proteins 
with no tails are shown in Fig. \ref{fig:figexnotails}. These were obtained by solving full PB equation, without 
linearizing it.
We see that the phase diagram in this case is totally ''flat'' - the smallest possible ${\cal T}$-number structure will always be 
the one with smallest complex energy, irrespectively of the core charge and its radius (at least in the range of values considered here, 
and the conditions summarized in \ref{subs:thermod}). 
Thus, the phase diagram is governed solely by simple geometry. The same result comes out also in the DH approximation. Note also how 
the assembly free energy shows practically no dependence on $\sigma_1$ and $R_1$ when $R_2 - R_1 \gg 1/\kappa$. This is due to 
the fact that in this regime, the assembly free energy is simply the (positive) electrostatic self-energy of the capsid which depends 
only on $R_2$, i.e. the ${\cal T}$ number of the capsid (see also Eq. (\ref{eq:deltfDH}) for the Debye-H\"{u}ckel description of this 
situation). When $R_2 - R_1 \sim 1/\kappa$ (i.e. about 1 nm in our case, $c_0=100$ mM), the electrostatic attraction between the core and the capsid 
becomes only partially screened by salt ions, and the free energy of the assembly, $\Delta F$, becomes negative suggesting that the 
assembly is thermodynamically preferrable, releasing extra energy in the solution. The assembly is also more efficient for larger core charge densities.

We still need to consider what happens in the case when the protein charges are delocalized on the flexible 
polyelectrolyte tails. In particular, the decrease of the volume of the space between the core and the capsid 
when $R_1 \rightarrow R_2$ would importantly confine the polyelectrolyte tails. We can thus expect 
their entropic and self-interaction contributions to become of the largest importance in the region where 
the assembly free energy of the tailless monomers shows the most negative values.

\subsection{Protein subunits with N-tails}
\label{sec:tails}

\begin{figure}[ht]
\centerline{
\epsfig {file=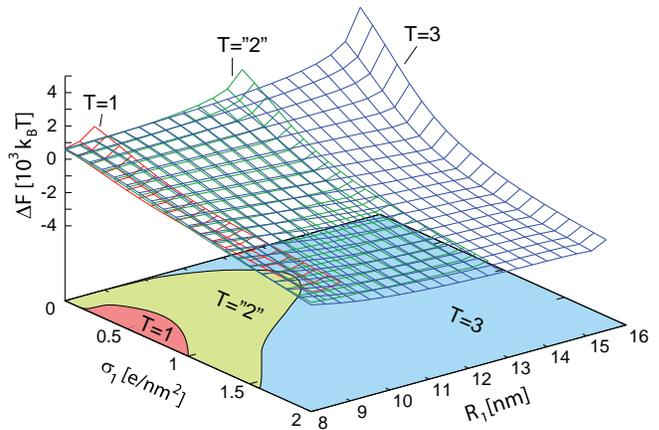,width=8.5cm}
}
\caption{Assembly free energies calculated as functions of the core radius $R_1$, its charge $\sigma_1$, for three 
different capsid radii, $R_2$ (and $N_p$), as quantified  by three different ${\cal T}$ numbers. The parameters of 
this calculation are $a=0.5$ nm, $p=1$, v=$0.5$ nm$^3$, $c_0=100$ mM, $N_t=18$. }
\label{fig:freentail}
\end{figure}

We now assume that the protein charge is distributed along the flexible N-tails, i.e. we take all the details of the 
model developed in Sec. \ref{sec:theory} into account. First, we calculate the free energies of the complex without the 
account of the finite extensibility of the tails, i.e. we solve Eqs. (\ref{eq:PBE}) and (\ref{eq:polyel}) with 
boundary conditions for the polyelectrolyte amplitude as specified in Eqs. (\ref{eq:impen1}) and (\ref{eq:anchor}). From the 
thus obtained free energy, we subtract the electrostatic self-energy of the core. The ensuing assembly free energies 
are shown in Fig. \ref{fig:freentail}.

We see that although the assembly free energies are of the same order of magnitude as in the case of tailless capsomeres, the 
shape of the free energy landscape is quite different and the assembly is now governed both by the core charge and by its radius. This is 
most easily seen by the complicated shape of the regions denoted by ${\cal T}=1$, ${\cal T}="2"$, and ${\cal T}=3$ in the $\sigma_1 - R_1$ plane in 
Fig. \ref{fig:freentail}, which correspond to capsids with the lowest 
free energy. We also see that when the two radii are close to each other, the free energy steeply rises. This is easy to understand, 
since in this case, the tails are forced to redistribute in a small volume in between the capsid and the core, so the contribution of 
entropic confinement to the free energy becomes significant. Interestingly, in the case of an infinitely thin capsid studied in the 
previous section, it was in these regions that the assembly free energy sharply {\em dropped}, exactly the 
opposite from the case we have here. It is thus clear at this point that the tails do introduce 
different physics into the problem. The assembly free energy in general also decreases with $\sigma_1$ which is an effect 
that we saw in the previous subsection and is due solely to electrostatics. We also observe that the values of the free 
energy are mostly smaller from the ones obtained in the model of an infinitely thin charged 
capsid, which essentially means that the tails 
can adopt such conformations that reduce the electrostatic part of the free energy to a significant extent, 
especially in the electrostatically unfavorable regime. Note also how the structure with the lowest assembly free energy increases its 
radius (and total charge) as the core charge increases (e.g. for $R$=8.25 nm one can see the progression from ${\cal T}=1$, ${\cal T}=2$ to 
${\cal T}=3$ structures as $\sigma_1$ increases). This can be simply explained by the screening of the core by the capsomer tails - the more charged 
the core, the more capsomeres (i.e. larger ${\cal T}$ numbers) are needed to screen it efficiently. But note here that the 
tails are assumed to be maximally flexible and can thus easily stretch from practically arbitrary distances (capsid) to the core in order 
to screen it. For given core radius, $R_1$, the assembly free energy is positive when $\sigma_1=0$, and it 
decreases as $\sigma_1$ increases, becoming negative for some ''critical'' core charge density, whose  
typical values in the range of parameters considered are $\sigma_1 \sim$ e/nm$^2$. Thus, in order for the assembly to proceed 
spontaneously, the cores need to be sufficiently charged.

\begin{figure}[ht]
\centerline{
\epsfig {file=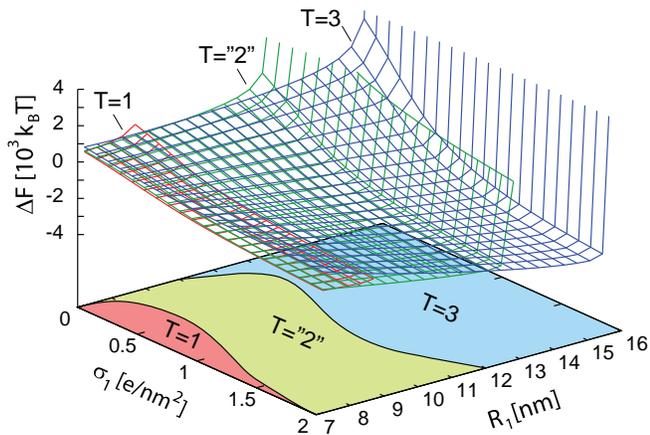,width=8.5cm}
}
\caption{Assembly free energies as functions of the core radius $R_1$, its charge $\sigma_1$, for three 
different capsid radii, $R_2$ (and $N_p$), as quantified  by three different T numbers. The maximal extensibility 
ansatz was used to obtain these results. The parameters of 
this calculation are $a=0.5$ nm, $p=1$, v=$0.5$ nm$^3$, $c_0=100$ mM, $N_t=18$. }
\label{fig:freentailansatz}
\end{figure}

\begin{figure}[ht]
\centerline{
\epsfig {file=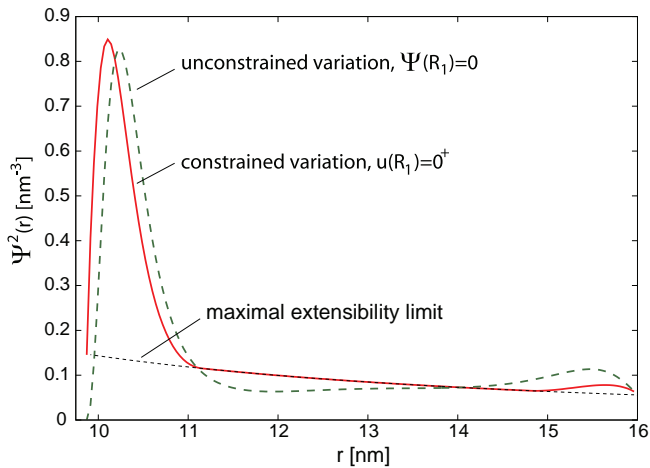,width=8.5cm}
}
\caption{Comparison of the unconstrained monomer density (thick dashed line) and constrained monomer density (full line) approaches. 
The parameters of the calculation are $\sigma_1=0.5$ e/nm$^2$, $c_0=100$ mM, $R_1=9.875$ nm, $R_2=16$ nm, $N_p=90$, $N_t=18$. 
The maximal extensibility limit [Eq.(\ref{eq:maxstretch})] is denoted by a thin dashed line.}
\label{fig:conc1}
\end{figure}

To see whether the results are influenced by the maximal extensibility of the tails, in Fig. \ref{fig:freentailansatz} we plot 
the assembly free energy using the maximal extensibility ansatz in Eq. (\ref{eq:maxstretch}) together with boundary 
conditions in Eq. (\ref{eq:ubound}). 
Again, the assembly free energies are of the same order of magnitude, but the borders between the different 
regions in the $\sigma_1 - R_1$ plane are quite different with respect to those obtained without the maximal 
extensibility ansatz. In order to better understand the results obtained thus far, it helps to plot the spatial 
distribution of the protein monomers, i.e. $\Psi^2(r)$, within the space between the core and the inner capsid radius. 
A solution for the case when  $N_p = 90$ (1620 monomers in total), $R_2 = 16$ nm (${\cal T}$=3), $R_1=9.875$ nm, 
$\sigma_1 = 0.5$ $e$/nm$^2$ is shown in Fig. \ref{fig:conc1}.

\begin{figure}[ht]
\centerline{
\epsfig {file=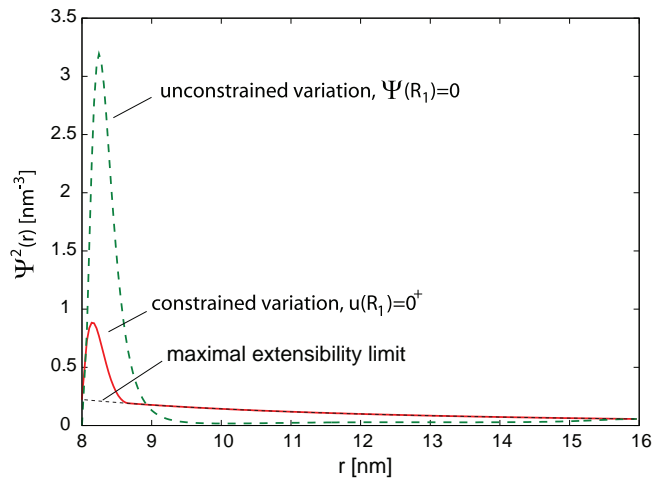,width=8.5cm}
}
\caption{Comparison of the unconstrained monomer density (thick dashed line) and constrained monomer density (full line) approaches. 
The parameters of the calculation are $\sigma_1=1.5$ e/nm$^2$, $c_0=100$ mM, $R_1=8$ nm, $R_2=16$ nm, $N_p=90$, $N_t=18$. 
The maximal extensibility limit [Eq.(\ref{eq:maxstretch})] is denoted by a thin dashed line.}
\label{fig:comp2}
\end{figure}

What this figure nicely illustrates is that the tails stretch out from the capsid surface towards the core so that they 
accumulate around the oppositely charged core. There are still some monomers in the space between 
the core and the capsid, but the dominant density is situated in a shell around the core. One can also see how the constrained 
solution stays above the maximal extensibility limit in Eq.(\ref{eq:maxstretch}). One should keep in mind that the boundary 
conditions for the two solutions are different, so one cannot expect that the constrained variation will always yield 
larger energy. This is also the case for the displayed calculation where we find $F=927$ $k_B T$, and $F=788$ $k_B T$ for 
the unconstrained and constrained values of the complex free energy, respectively. Note how the constrained polyelectrolyte density in fact
approaches closer to the charged core than the unconstrained density due to the different boundary conditions that the two 
satisfy. The noted effect results in the lowering of the free energy in the constrained case. However, this is not always the case, and 
it depends on the distance between the core and the capsid ($R_2 - R_1$). When this distance is sufficiently large, the tails cannot 
accumulate around the core even when they maximally stretch, so that they cannot screen the core efficiently. This effect is not 
present in the calculation with the unconstrained polyelectrolyte amplitude. This is in fact the most important reason for the 
different look of boundaries in the $\sigma_1 - R_1$ plane for the two calculations. Note how in the unconstrained case, the 
complex of core with ${\cal T}=3$ capsid has the lowest energy for sufficiently large $\sigma_1$ ($\sigma_1>1.5$ e/nm$^2$) 
in a huge range of radii $R_1$ (8-16 nm, see Fig. \ref{fig:freentail}). However, the maximal length of the (fully stretched) 
tails is $N_t a = 9$ nm, so that for $R_1=8$ nm tails are very much extended and only a small part of the polyelectrolyte 
density can gather around the core. As the unconstrained results do not account for this effect, by breaking the maximal 
extensibility limit in Eq. (\ref{eq:maxstretch}) the polyelectrolyte density screens the core efficiently and thus 
lowers the free energy of the ${\cal T}=3$ structure with respect to the value obtained in the constrained calculation. 
This effect is illustrated in Fig. \ref{fig:comp2} for ${\cal T}=3$ ($R_2=16$ nm, $N_p=90$), $R_1=8$ nm, $\sigma_1=1.5$ e/nm$^2$. Note 
how the maximal extensibility limit is severely broken in the unconstrained calculation.
The free energies of the complexes are $F=2710$ $k_B T$, and $F=3520$ $k_B T$ in the unconstrained and constrained calculations, respectively, 
demonstrating the effect discussed.

\begin{figure}[ht]
\centerline{
\epsfig {file=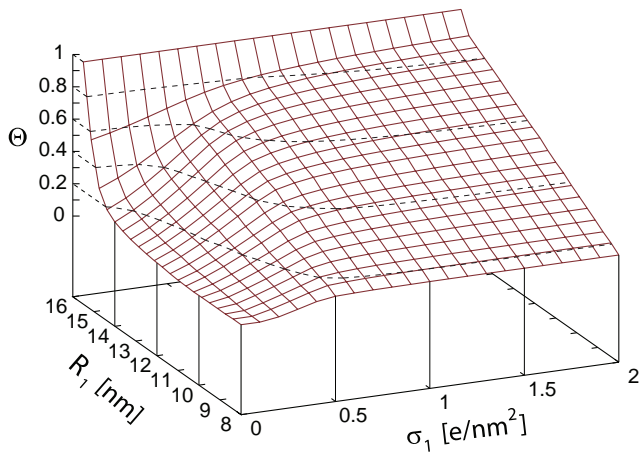,width=8.5cm}
}
\caption{Coverage of the core defined by Eq. (\ref{eq:thetacov}) as a function of $\sigma_1$ and $R_1$ for a complex 
with ${\cal T}$=3 capsid ($R_2=16$ nm). The calculation was performed with the maximal extensibility constraint on the polyelectrolyte 
amplitude.
}
\label{fig:cover}
\end{figure}

It is of interest to examine the efficiency of core screening by the polyelectrolyte tails somewhat closer. We define the ratio 
\begin{equation}
\Theta = \frac{\int_{R_1}^{R_1+a} d^3 r \Psi^2(r)}{N_t N_p},
\label{eq:thetacov}
\end{equation}
which can be interpreted as a percentage of monomers in a shell of thickness $a$ (monomer-monomer separation) around the core. 
This can also be thought of as the percentage of monomers that cover (are ''in contact with'') the core. In Fig. \ref{fig:cover} 
we display the coverage $\Theta$ as a function of $\sigma_1$ and $R_1$ for ${\cal T}=3$ capsid ($R_2=16$ nm).

What can be easily seen from this plot is the gradual shift of the polyelectrolyte density from the space close to capsid to 
the shell surrounding the core as the core charge density ($\sigma_1$) increases. Due to the maximal extensibility constraint that was 
included in these calculations, $\Theta$ parameter saturates at a value smaller than 1 as $\sigma_1$ increases. This simply 
reflects the fact that the polyelectrolyte must pass through the region between the core and the capsid in order to 
accumulate around the core, and so much of its density may remain in the space in between the capsid and the core. This 
effect becomes more important when $(R_2 - R_1)$ becomes comparable to the tail length, $N_t a$. The charge density of the core at which 
one observes the significant accumulation of the monomers around the core ($\sigma_1 \sim 0.5$ e/nm$^2$) is the same as 
the charge density at which the assembly free energy becomes negative, i.e. the assembly proceeds spontaneously.

\section{Conclusions}

\begin{figure}[ht]
\centerline{
\epsfig {file=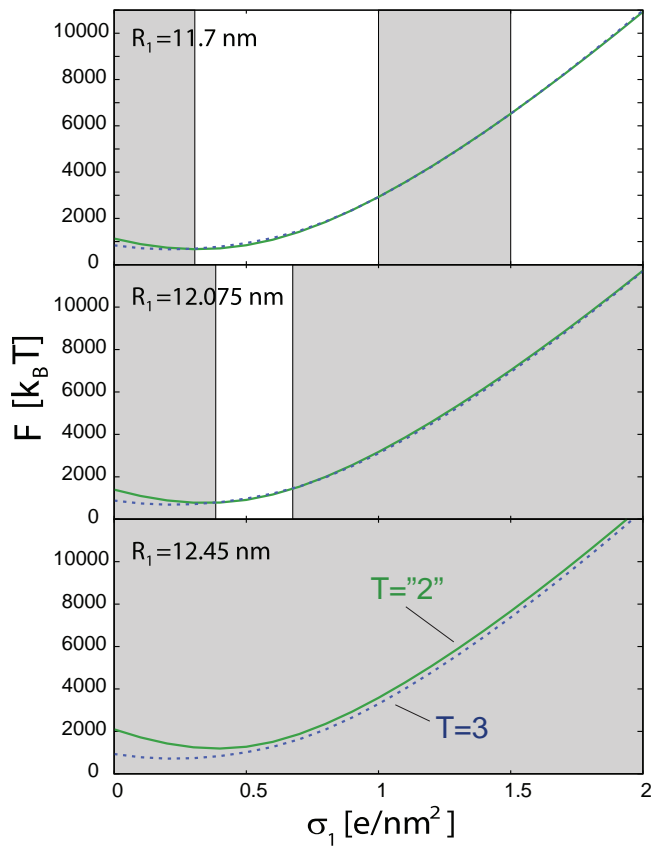,width=8.5cm}
}
\caption{Free energy of the complexes for three different core radii in the transition region as a function of core 
surface charge density. The dotted and full lines display the results for ${\cal T}$=3 and "2" complexes, respectively. 
The parameters of the polyelectrolyte are the same as before and the calculations were performed for the constrained 
polyelectrolyte model (maximal extensibility limit obeyed). Grey regions denote the surface charge densities $\sigma_1$ 
where the ${\cal T}$=3 structure has lower free energy.
}
\label{fig:transit}
\end{figure}

An intriguing feature of our results is that both the core charge and its radius determine the size 
of the capsid around the core. A particularly interesting case is when the core radius is close (but somewhat 
smaller) to the T=1 capsid, {\it i.e.}, $R_1 = 8$ nm. For sufficiently small core surface charge density ($\sigma_1 < 0.5$ e/nm$^2$), 
${\cal T}=1$ structures around the cores shall form. However, if the surface charge density increases over some critical value 
(around 1.0 e/nm$^2$ in the constrained model of the polyelectrolyte), ${\cal T}$="2" capsids shall form, in spite of the fact 
that the core radius is more than 5 nm smaller from the radius of ${\cal T}$="2" capsid (see Fig. \ref{fig:freentailansatz}). 
This clearly shows that in addition to core radius, which dominantly influences the assembly process, one needs to have an adequate 
charge density in order to produce the structures of desired T-number. 
The same effect is present on the ${\cal T}$="2" and ${\cal T}=3$ border when $R_1 > 11.6$ nm. We show this transition region 
in another way in Fig. \ref{fig:transit}.
Note that although the free energy curves indeed cross for certain values of $\sigma_1$ and $R_1$, their magnitudes remain 
quite similar even deeply in the transition region. Thus one could expect to observe a polydisperse distribution of ${\cal T}$="2" 
and ${\cal T}=3$ structures in a solution with a monodisperse distribution of core size and charge density. 

An interesting 
effect is observed for $R_1=12.075$ nm in Fig. \ref{fig:transit} where there exists {\em two} intersections between 
the free energy curves for ${\cal T}$="2" and ${\cal T}$=3 complexes as $\sigma_1$ increases. If the charge density is 
lower than 0.3, ${\cal T}$=3 structures have the lowest assembly free energy. Quite surprisingly, upon increasing the charge density, 
${\cal T}=2$ structures become the dominant ones, i.e. the size of thermodynamicaly preferred capsids decreases. This is mainly due to the fact 
that with increasing the charge density, the monomers in the N-terminal tails prefer to sit next to the core,i.e., the electrostatic 
interaction wins over the chain configurational entropy.  However, as we increase the core charge density beyond 0.7, it becomes more 
advantageous to have $T=3$ structures again as there will be more charges associated with the N-tails of ${\cal T}$=3 structures.

An even more interesting effect is illustrated in Fig. 9 for $R_1 = 11.7$ cores. Here, we observe three transition lines instead of two 
of the previous case. Somewhat counter-intuitively, for the charge densities above 1.3, ${\cal T}$="2" particles become free energy minima structures again.
This indicates that under physiological conditions, if we increase the charges on the cores significantly, the lowest possible ${\cal T}$ structures (${\cal T}=2$ in 
this case) form so that more charges on the N-tails can sit in the immediate vicinity of the core. This is due to the maximal extensibility constraint, and the 
effect is not present in the calculation with the tails that do not satisfy the constraint (see Fig. \ref{fig:freentail}). For sufficiently charged cores 
($\sigma_1>2$ e/nm$^2$), it turns out to be better again to have smaller number of total charges (i.e. the ${\cal T}=2$ instead of ${\cal T}=3$ capsid) that can 
approach the capsid easily without any geometrical constraint.    
 
It is obvious from the previous discussion that the finite extensibility of the 
tails is important for determination of the lowest energy structures. One can see this most easily by comparing Figs. 
\ref{fig:freentail} and \ref{fig:freentailansatz}. This effect becomes particularly important for structures in which the 
core is significantly smaller than the capsid, i.e. when $(R_2 - R_1)/(N_t a) \gtrsim 1$. 

\begin{figure}[ht]
\centerline{
\epsfig {file=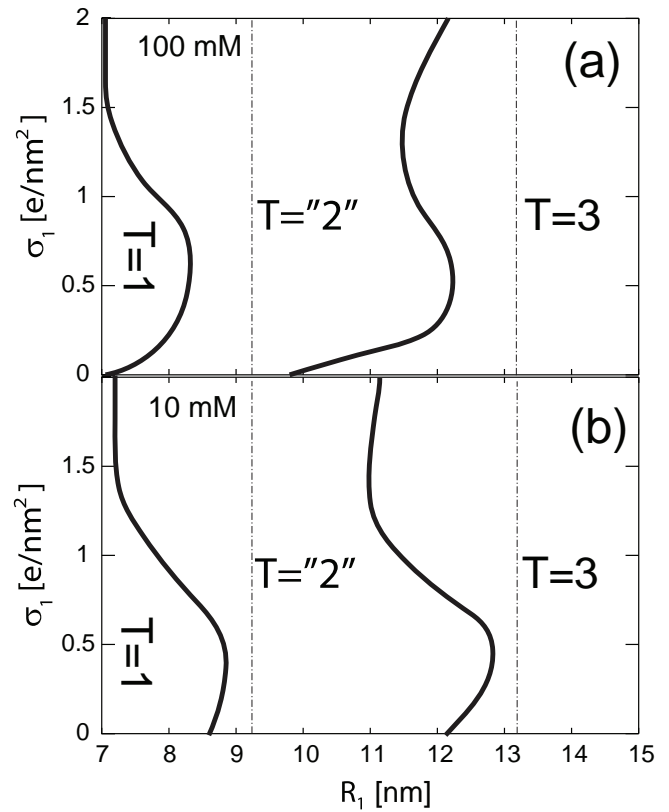,width=8.5cm}
}
\caption{Regions in the $R_1 - \sigma_1$ space in which a particular ${\cal T}$-number structure has the lowest free energy, 
as denoted. Panels (a) and (b) show the results for monovalent concentrations of $c_0=100$ mM and $c_0=10$ mM, respectively. 
The two vertical dash-dotted lines denote the radii of ${\cal T}=$1, and "2" structures ($R_2$).}
\label{fig:figphase}
\end{figure}

To test the theory of our previous paragraph , we repeated our calculations at lower salt concentrations.  The results are presented in Fig. \ref{fig:figphase} 
As it is shown in the figure the transition line from ${\cal T}$=3 to ${\cal T}$="2" is moved to very high $\sigma$ as expected. 


In summary, we have demonstrated that the thermodynamics of the assembly nontrivially depends on the 
electrostatic and geometric constraints which include $\kappa(R_1 - R_2)$ (electrostatic screening),  maximal possible 
stretching $(R_2 - R_1)/(N_t a) \sim 1$, and confinement $(R_2 - R_1)/ a < 1$ of the protein tails. Our study provides 
quantitative guidelines for experiments aiming to assemble ''hybrid'' structures, i.e. protein shells around 
charged and impenetrable cores, especially in the limit when the number of cores is much smaller than the number of 
proteins.

\section*{Acknowledgments}

The authors would like to acknowledge helpful discussions with B. Dragnea.  A. \v{S}. acknowledges support by the Ministry of Science, Education, and Sports of Republic of Croatia 
(Project No. 035-0352828-2837).   R.P. acknowledges the financial support by 
the Slovenian Research Agency under contract No. P1- 0055 (Biophysics of Polymers, Membranes, Gels, Colloids 
and Cells) and J1-0908 (Active media nanoactuators with dispersion forces). This study was 
supported in part by the Intramural Research Program of the NIH, Eunice Kennedy Shriver National Institute of Child Health (R.P.) and NSF grant DMR-06-45668 (R.Z.).

\section*{Appendix: Derivation of Debye-H\"{u}ckel formulas}

We solve the linearized Poisson-Boltzman equation for the system with tailless capsomeres. 
The electrostatic potential can be written in region I (see Fig. \ref{fig:ilustracija}) as
\begin{equation}
\Phi_{I}(r) = C,
\end{equation}
in region II as 
\begin{equation}
\Phi_{II}(r) = A \frac{\exp (-\kappa) r}{r} + B \frac{\exp(\kappa r)}{r},
\end{equation}
and in region III as
\begin{equation}
\Phi_{III}(r) = D \frac{\exp (-\kappa r)}{r},
\end{equation}
where $\kappa$ is the Debye-H\"{u}ckel screening length, and $A,B,C$, and $D$ are unknown constants that 
are to be determined from the two boundary conditions at $R_1$ and two at $R_2$. This yields
\begin{equation}
A = \frac{e^{\kappa(R_1-R_2)} \left[ 2 \kappa R_1^2 \sigma_1 e^{\kappa R_2} + 
(\kappa R_1-1)R_2 \sigma_2 e^{\kappa R_1}\right]}
{2 \epsilon_0 \epsilon_r \kappa (1+\kappa R_1)},
\end{equation}
\begin{equation}
B = \frac{R_2 \sigma_2 e^{-\kappa R_2}}{2 \epsilon_0 \epsilon_r \kappa},
\end{equation}
\begin{equation}
C = \frac{R_1 \sigma_1 + R_2 \sigma_2 e^{\kappa(R_1-R_2)}}{\epsilon_0 \epsilon_r (1 + \kappa R_1)},
\end{equation}
and
\begin{eqnarray}
D = \frac{e^{-\kappa R_2}}{2 \epsilon_0 \epsilon_r \kappa (1 + \kappa R_1)} &\times& \left[ 
 2 \kappa R_1^2 \sigma_1 e^{\kappa(R_1+R_2)} \right. \nonumber \\
&+& \left. (\kappa R_1 - 1) R_2 \sigma_2 e^{2 \kappa R_1} \right. \nonumber \\
&+& \left. (1+\kappa R_1) R_2 \sigma_2 e^{2 \kappa R_2} \right].
\end{eqnarray}
The electrostatic free energy can be written as
\begin{equation}
F = \int d^3 r \frac{Q(r) \Phi(r)}{2},
\end{equation}
which yields
\begin{eqnarray}
F &=& \frac{\pi e^{-2 \kappa R_2}}{\epsilon_0 \epsilon_r \kappa (1+\kappa R_1)} \times \left \{
4 \kappa R_1^2 R_2 \sigma_1 \sigma_2 e^{\kappa(R_1+R_2)} \right. \nonumber \\
&+& (\kappa R_1-1) R_2^2 \sigma_2^2 e^{2 \kappa R_1} \nonumber \\
&+& \left. [2\kappa R_1^3 \sigma_1^2 + (1+\kappa R_1) R_2^2 \sigma_2^2 ] e^{2 \kappa R_2} \right \}
\label{eq:horrible}
\end{eqnarray}

\end{document}